\documentclass[preprint,aps,twocolumn,10pt,final]{revtex4}
\input{tcilatex}

\begin{document}

\title{Experimental Realization of the Quantum Box Problem}
\author{K.J. Resch$^{1}$, J.S. Lundeen$^{2}$, and A.M. Steinberg$^{1,2}$}
\affiliation{$^{1}$Institut f\"{u}r Experimentalphysik, Universit\"{a}t Wien,
Boltzmanngasse 5, 1090, Vienna, AUSTRIA\\
$^{2}$Department of Physics, University of Toronto,\\
60 St. George Street, Toronto, ON M5S 1A7 CANADA}

\begin{abstract}
The three-box problem is a \emph{gedankenexperiment} designed to elucidate
some interesting features of quantum measurement and locality. A particle is
prepared in a particular superposition of three boxes, and later found in a
different (but nonorthogonal) superposition. It was predicted that
appropriate ``weak'' measurements of particle position in the interval
between preparation and post-selection would find the particle in two
different places, each with certainty. We verify these predictions in an
optical experiment and address the issues of locality and of negative
probability.
\end{abstract}

\maketitle

\newpage

Weak measurements have been controversial ever since the concept was
developed by Aharonov, Albert, and Vaidman (AAV) \cite{weakmeas}. \ In
contrast to the usual, von Neumann, approach to measurement, weak
measurement uses an apparatus whose pointer has a very large quantum
mechanical uncertainty when compared with its typical shift. \ After the
system-pointer interaction, the shift in the pointer position is much
smaller than its initial uncertainty and almost no information is gained
about the quantum system. \ Nevertheless, after a sufficiently large number
of measurements on an ensemble of identically prepared quantum systems, the
mean pointer position can be determined to any degree of precision. \ In
such a measurement strategy, one sacrifices knowledge of the value of an
observable on any given experimental run to avoid entanglement with the
measurement device and the ensuing `collapse' of the wavefunction. \ In
particular, this makes it possible to contemplate the behavior of a system
defined both by state preparation and by a later post-selection, without
significant disturbance of the system in the intervening period.

AAV \cite{weakmeas} calculated the shift in the pointer of a measurement
apparatus that weakly measured an observable $A$ between two strong
measurements. \ The initial strong measurement pre-selects (or prepares) the
state, $\left| \psi _{i}\right\rangle ,$ of the quantum system and the final
strong measurement post-selects the quantum state, $\left| \psi
_{f}\right\rangle $. \ In between, consider a von Neumann-style interaction
Hamiltonian of the form%
\begin{equation}
\mathcal{H}_{I}=g\hat{A}\hat{P_{x}},  \label{interaction}
\end{equation}%
where $\hat{A}$ is the hermitian operator corresponding to an observable $A$
of the quantum system, $g$ is a (real) coupling constant, and $\hat{P}_{x}$
is the momentum operator conjugate to the pointer position $\hat{X}.$\ In
the absence of postselection, the effect of having this measurement
interaction on for a time $T$ (assumed short enough that $A$ is constant
during the measurement) shifts the pointer position by an amount $\Delta
x=K\left\langle \hat{A}\right\rangle \equiv gT\left\langle \hat{A}%
\right\rangle ,$such that one can infer a value for $A$ by dividing the
pointer shift by the interaction strength $K$. \ The main result of AAV's
seminal work \cite{weakmeas} is that for sufficiently weak coupling strength 
$K$ and in the presence of postselection, the inferred value of $A$ is given
by the following expression, which they called the ``weak value'':%
\begin{equation}
A_{W}=\frac{\left\langle \psi _{f}\right| \hat{A}\left| \psi
_{i}\right\rangle }{\left\langle \psi _{f}|\psi _{i}\right\rangle }.
\label{weakform}
\end{equation}%
When the two strong measurements pre- and post-select the same state, $A_{W}$
reduces to the usual quantum mechanical expectation value for the operator $%
\hat{A}$. \ However when the pre- and post-selected states differ, weak
values can take on surprising values that are not constrained to lie within
the eigenvalue spectrum of the operator or even to be real numbers. \ The
surprising character of weak values has led to skepticism about whether they
should be considered proper measurements \cite{leggett,peres,reply}. \ In
spite of their controversial nature, a weak measurement has been
experimentally performed by Ritchie \emph{et al.} \cite{weakexp}; in
addition, they have been useful in correctly describing or predicting
surprising experimental outcomes \cite%
{weakmeas,opticalweak,aetunnelweak,hardyvaid,wiseman,molmer,orozco}.

The ``Quantum Box Problem'' was developed by Aharonov and Vaidman \cite%
{box,weakelements}; it is a deceptively simple thought experiment that
elucidates some of the odd behavior which may result from studying
post-selected systems. \ In the 3-box version, a particle is prepared
(pre-selected) in an equally weighted superposition of being in one of three
orthogonal quantum ``boxes'', A, B, and C, i.e., $\left| \psi
_{i}\right\rangle =\frac{1}{\sqrt{3}}\left| A\right\rangle +\frac{1}{\sqrt{3}%
}\left| B\right\rangle +\frac{1}{\sqrt{3}}\left| C\right\rangle $ and
post-selected in the final state $\left| \psi _{f}\right\rangle =\frac{1}{%
\sqrt{3}}\left| A\right\rangle +\frac{1}{\sqrt{3}}\left| B\right\rangle -%
\frac{1}{\sqrt{3}}\left| C\right\rangle $. \ The weak value for the particle
being in box A can be calculated by inserting the projection operator $%
\left| A\right\rangle \left\langle A\right| $ in Eq. \ref{weakform}, i.e., $%
A_{W}=\left\langle \psi _{f}|A\right\rangle \left\langle A|\psi
_{i}\right\rangle /\left\langle \psi _{f}|\psi _{i}\right\rangle =+1$. \ In
this case, the pointer shift is the same as in the case when the particle
was definitely in box A. \ We call this quantity the ``weak probability for
the particle in box A'' and use the shorthand $P_{AW},$ where $A$ represents
the rail. This weak probability is already strange since a strong
measurement would find the particle in box A only $\left| 1/\sqrt{3}\right|
^{2}=1/3$ of the time. \ Through a similar calculation, the weak probability
for the particle in box B, $P_{BW}$, is also found be be $+1$. \ Like normal
probabilities, the sum of all weak probabilities is $+1$ (because the sum of
all orthogonal projectors is the identity); therefore, if one performs a
weak measurement of the particle in box C one finds the weak probability, $%
P_{CW}=$\ $\left\langle \psi _{f}|C\right\rangle \left\langle C|\psi
_{i}\right\rangle /\left\langle \psi _{f}|\psi _{i}\right\rangle =-1$. \
Such a result is very strange since it lies outside the range of eigenvalues
for projection operators and normal probabilities -- i.e., it does not lie
between $0$ and $1$. \ Nevertheless, for a large enough ensemble of
identically prepared and post-selected states, this result properly predicts
the outcome of our experiment. \ Note that while apparently similar to the
situation for quasi-probability distributions such as the Wigner function,
this case is different, referring instead to a directly observable
measurement outcome. \ There are also other situations in which it has been
suggested that negative probabilities might be useful for resolving locality
``paradoxes'' \cite{negprobs}. \ For a review of such extensions to
probability theory, see \cite{muckenheim}.

We use a linear optical interferometer to implement the quantum box problem
(Fig. 1). \ This interferometer is similar to the Mach-Zehnder
interferometer except that it has three optical paths (or rails) instead of
two; we label these rails A, B, and C. \ A photon can be prepared in any
superposition of these three rails by proper selection of the
characteristics of beamsplitters BS1 and BS2. \ Similarly, any coherent
superposition can be post-selected by controlling the characteristics of
BS3, BS4, and the optical path lengths (relative phases). \ We use the
transverse displacement of the photon as our measurement pointer. \
Measurements are carried out by tilting one of the glass plates (GP A, B, or
C); this results in a controllable transverse shift of the beam in one of
the rails. \ For example, a glass plate in rail A displacing the beam along $%
X$ can be thought of in terms of an effective interaction Hamiltonian of the
form $g\left| A\right\rangle \left\langle A\right| \hat{P}_{X}$. \ The
measurement of the mean spatial shift of an ensemble of photons can be
viewed as a measurement of the probability for each photon to have been in
rail A. \ If the transverse shift is much less than the beam size (the
uncertainty in the position of the individual photons), this is a weak
measurement. \ Shifting only one rail at a time, we measure the size of that
shift while keeping the other two rails blocked. \ This characterizes the
strength $K_{A}$ of the system-pointer interaction. \ We then combine all
three beams with the correct relative phases for proper post-selection and
again measure the shift $\Delta x$. \ The ratio of the shift in the
post-selected state to that of the single rail constitutes our weak
probability for that rail $A_{W}=\Delta x/K_{A}$. \ A negative weak
probability is realized when the shift in the post-selected state is in the 
\emph{opposite direction} to the shift on the individual rail.

For the experiment, light from a 780nm, 30mW, diode laser is spatially
filtered to yield a collimated TEM$_{00}$ gaussian beam with a waist of 380$%
\mu $m; this amounts to a large ensemble of identically-prepared photons. \
To create the appropriate superposition, the transmissivity and reflectivity
of each beam-splitter is adjusted by using two half-wave plates ($\lambda /2$%
) prior to polarizing beam-splitters (PBS). \ The third half-wave plate is
present to rotate the polarization in rail C from horizontal to vertical so
that all three rails have the same polarization. \ Glass plates with
thicknesses 10mm, 6.5mm, and 10mm are situated in paths A, B, and C
respectively. \ The plate in rail A displaces the beam in the vertical
direction (out of the plane of the inteferometer) while the plates in rails
B and C displace the beam in the horizontal direction (in the plane of the
interferometer). \ In addition to these glass plates, 1-mm-thick microscope
slides are located in rails A and C. \ These plates can also be finely
tilted to set the relative phases, $\phi _{A}$ and $\phi _{C}$, between the
light in these arms with that in rail B without significantly displacing the
beam. \ Rails A and B are recombined coherently at beam-splitter (BS) 3. \
One of the outputs from BS3 is then recombined with the light from rail C at
BS4. \ The phases are chosen such that the light in the right-going output
of BS4 is our desired post-selected state. \ This beam is magnified by a
factor of 3 on a screen behind which a CMOS\ camera (Logitech Quickcam)
captures beam images. \ The second output of BS4 terminates at a photodiode
that is used to set the relative phases of the three arms of the
interferometer.

The quantum box problem can be generalized for an arbitrary pre-selected
state $\left| \varphi _{i}\right\rangle =a_{i}\left| A\right\rangle
+b_{i}\left| B\right\rangle +c_{i}\left| C\right\rangle $ and post-selected
state $\left| \varphi _{f}\right\rangle =a_{f}\left| A\right\rangle
+b_{f}\left| B\right\rangle -c_{f}\left| C\right\rangle .$ \ If the
coefficients are real numbers, we obtain the same weak values discussed
earlier for rails A, B, and C if $a_{i}a_{f}=b_{i}b_{f}=c_{i}c_{f}$, albeit
with a lower overlap between our pre- and post-selected states. \ These
requirements can be converted into experimental parameters in the following
way. \ Given reflection and transmission amplitudes $r_{i}\exp (i\phi _{ri})$
and $t_{i}\exp (i\phi _{ti})$ for BS$i$, the 3-path interferometer
pre-selects the state $\left| \varphi _{i}\right\rangle =r_{1}\left|
A\right\rangle +t_{1}r_{2}\left| B\right\rangle +t_{1}t_{2}\left|
C\right\rangle $ -- the phases are compensated for by optical path lengths$.$
\ For the proper phase settings, the interferometer post-selects the state $%
\left| \varphi _{f}\right\rangle =t_{3}r_{4}\left| A\right\rangle
+r_{3}r_{4}\left| B\right\rangle -t_{4}\left| C\right\rangle .$ To satisfy
the condition $a_{i}a_{f}=b_{i}b_{f}=c_{i}c_{f}$, we require $%
r_{1}t_{3}r_{4}=t_{1}r_{2}r_{3}r_{4}=t_{1}t_{2}t_{4}$. \ This is the
condition that each path contributes the same intensity in the camera output
port. \ In our experiment, the final beamsplitters are both 50/50, and
therefore our post-selected state is approximately $\left| \varphi
_{f}\right\rangle =\frac{1}{2}\left| A\right\rangle +\frac{1}{2}\left|
B\right\rangle -\frac{1}{\sqrt{2}}\left| C\right\rangle .$ \ The proper
pre-selected state to obtain the desired weak values is $\left| \varphi
_{i}\right\rangle =\sqrt{\frac{2}{5}}\left| A\right\rangle +\sqrt{\frac{2}{5}%
}\left| B\right\rangle +\sqrt{\frac{1}{5}}\left| C\right\rangle .$ \ We note
that this reduces the overlap of the initial and final states from $1/3$ for
the original states to $\sqrt{1/10}.$

The data for this experiment were taken in two parts. \ In both parts, we
began by balancing the intensities in the camera arm from the three paths
and aligned the beams in the interferometers such that they overlapped to
better than 1/10 of their rms\ widths. \ All of the beams in the
interferometer were vertically-polarized. \ Pairwise, beams A\&B and B\&C
had interference fringe visibilities of about 95\%. \ In the first part of
the experiment we performed only single weak measurements on each of the 3
rails. \ A single glass plate (GP) in one of the arms was tilted to displace
the beam in that path. \ The phases in the interferometers were set using
the microscope slides so that paths A and B interfere constructively and
paths B and C\ interfere destructively in the camera output. \ Images of
each individual beam (in the absence of interference) and the properly
post-selected beam were recorded using the camera. \ In Fig. 2, horizontal
profiles are shown, with a particular displacement of rail C, for the beams
in rails A (thin solid line), B (thin dashed line), and C (thick dashed
line) alone, and also for the post-selected state (thick solid line). \
Beams A and B have the same average position to within less than 1 pixel, or
about 1/20 of the rms\ width. \ Beam C has been displaced from this centre
position by $-11.1$ pixels or $-0.69$ rms widths. \ This is the coupling
strength $K_{C}$. \ The post-selected state beam profile is displaced by $%
+7.0$ pixels or $+0.44$ rms widths in the \emph{opposite} direction of rail
C; in other words, the observed value $\Delta x/K_{C}=-0.64$. \ This differs
from the expected value $P_{CW}=-1$ because the displacement is already
approaching the transition from the weak to the strong measurement regime.

We have summarized our results for post-selected measurements of $P_{A}$, $%
P_{B}$, and $P_{C}$ for a variety of coupling strengths in Fig. 3. \ The
observed displacements $\Delta x$ for measurements of $P_{A}$ and $P_{B}$
are shown as open circles and solid triangles respectively, as a function of 
$K_{\left\{ A,B\right\} }$. \ The theoretical prediction for these two rails
is a straight line with a slope of $1$ and a y-intercept of $0.$ \ The slope
of $1$ indicates a shift as large as if all the photons had traversed the
shifted rail, that is, $P_{AW}=P_{BW}=1$. \ Note that this slope of 1
persists even into the strong-measurement regime, $K_{\left\{ A,B\right\}
}>1 $ rms width$.$ \ This can be understood from the following argument. \
If we perform a weak measurement on rail A, then in our post-selection
output the amplitudes from rails B and C\ interfere perfectly destructively
and we are left only with the field from rail A. \ Therefore, whatever
displacement rail A has, so will the post-selected state \cite{weakelements}%
. \ Of course a similar argument can be constructed for rail B. \ The data
for rail C has different behaviour and no such argument applies. \ The
experimental data is shown as solid circles and the theoretical prediction
as a solid line. \ Near zero displacement of the rail, the displacement of
the post-selected state is in the opposite direction as the displacement of
rail C. \ The weak probabilities are given by the slopes of the curves from
the plot in Fig 3 near the origin. The weak probabilities are consistent
with the predictions of $+1$, $+1$, and $-1$ for $P_{AW}$, $P_{BW}$, and $%
P_{CW}$ respectively. \ For larger displacements in rail C, the weakness
criterion is not satisfied, and the observed shifts eventually approach a
strong-measurement limit of $1/5$ \cite{ABL}.

We have shown that the weak probability to find a particle in rail $A$ is $%
+1 $, as is the the weak probability to find the particle in rail $B$. This
begs the question, what is the weak probability to find the particle in both
arms? \ It is well-known that a single particle cannot trigger two different
spacelike separated detectors; in fact one might think of that as the
defining quality of a particle -- it must be either here or there but never
both. Therefore, one would never expect to find a single photon in both
rails A and B via projective measurements. \ It has been suggested, however,
that this is a limitation of strong measurements; if a particle is prepared
in a suitably delocalized state, might it be possible to find that particle
in two different places at the same time if the presence of the particle is
probed weakly \cite{fallingtree}?

For a state involving only one particle, we can use the operator $\left|
A\right\rangle \left\langle A|B\right\rangle \left\langle B\right| $ to find
the joint weak probability for the particle to be in rails A and B
simultaneously. \ However, because $\left| A\right\rangle $ and $\left|
B\right\rangle $ represent orthogonal states, this operator is identically
zero. \ Therefore the weak probability to find the particle in both rails is
expected to be zero even when the single weak probabilities (i.e., $P_{AW}$
and $P_{BW}$) are both equal to 1. \ This seeming violation of standard
rules of probability can be seen to be related to the relaxation of the
non-negativity axiom of probability theory \cite{hardyvaid,aewheeler}.

In the second part of the experiment, we performed two different
simultaneous measurements on the photons using two different degrees of
freedom at the same time. \ In addition to the transverse beam displacement
in rail B, we used the waveplate in rail C (see Fig. 1) just before BS4 to
rotate the polarization of the light by a small amount. We prepared the same
initial state as before but use a different final state, $\left| \psi
_{f2}\right\rangle =-\frac{1}{2}\left| A\right\rangle +\frac{1}{2}\left|
B\right\rangle +\frac{1}{\sqrt{2}}\left| C\right\rangle $. Such a final
state swaps the roles of rails A and C so that $P_{AW}=-1$ and $P_{CW}=+1$.
\ The reason for doing so was purely technical. Our final beamsplitters were
50/50 for vertically-polarized light but for horizontally-polarized light
they had greater than 90\% transmission. We ensured that rail C, where the
polarization rotation occurs, was transmitted rather than reflected, so that
the observable polarization shift at the camera would not be negligibly
small. \ We rotated the waveplate in rail C by a small amount that resulted
in a $K_{C}=9.6^{\circ }$ rotation in the polarization of the light \emph{%
after} the final beamsplitter when rails $A$ and $B$ were blocked. \ 

We measured the displacement, $\Delta x$, of the post-selected state as a
function of the rail-B coupling strength $K_{B}$. \ The results, $\Delta
x/K_{B}$, are the inferred probabilities of the photon to be in rail B and
are shown in Fig. 4., both in the case of no final polarizer (solid circles,
solid line theory) and in the case of a final polarizer oriented to block
vertically-polarized light (open circles, long-dashed line theory). \ Recall
that all of the beams in the interferometer initially had
vertically-polarized light. \ When the polarizer was inserted to block
vertically-polarized light we measured the beam displacement of only the
light that had its polarization rotated, i.e., only photons which had
traversed rail C. \ We also show the inferred probability of the photon to
be in rail C -- given by the polarization rotation of the final state
divided by $K_{C}=9.6^{\circ }$ (open triangles, dotted line theory). \ The
low data point for the $P_{B}$ was most likely due to a mode-hop-induced
phase fluctuation, and the slight disagreement between the theory and
experimental values for $P_{C}$ is from imperfect interference. \ These data
show, that for displacements of rail B in the weak regime (less than 0.5 rms
widths), the entire distribution of the light is both displaced and
polarization rotated as if the photons really experience both weak
measurements. \ However, by blocking all of the vertically-polarized light,
we clearly see that the distribution is not shifted at all. \ Therefore we
can say that those particles that were \emph{definitely} in rail C were not
even weakly in rail B. \ \ \ 

We have implemented the quantum box problem in a 3-rail\ interferometer and
verified some of the important predictions about the weak measurements one
can perform in this system. \ Specifically, we have observed weak
probabilities of $+1$ for rails A and B and a weak probability of $-1$ for
rail C. \ We have also studied the transition out of the weak-measurement
regime for rail C, when the interaction shifts the measurement pointer by
more than about 0.5 rms widths. \ In addition, we performed two simultaneous
measurements on two of the rails using polarization and transverse
displacement as pointers. \ For small displacements, we found that the
entire beam was shifted and polarization rotated. \ However, when only the
polarization-rotated light was studied, its transverse displacement was
zero. \ This shows the (one-sided) anticorrelation between a strong
measurement of the particle in one rail and a weak measurement of it in
another. \ This leaves open the question of what one would observe by making 
\emph{joint weak measurements} on two different rails. \ In another work, we
have proposed an experimentally feasible method for performing such
measurements \cite{jwv}, and an ongoing experiment aims to carry them out
for the 3-box problem. \ These types of conceptually simple experiments
yield insight into the sometimes shocking behavior of post-selected
subensembles in quantum mechanics, and into the power and the limitations of
weak measurements.

We would like to thank Morgan Mitchell, Jeff Tollaksen, and Jeeva Anandan
for helpful advice and discussions. \ We gratefully acknowledge financial
support from Photonics Research Ontario and NSERC.

\pagebreak

\bigskip

\textbf{Figure 1. \ The 3-rail Mach-Zehnder-style interferometer. \ The TEM}$%
_{00}$\textbf{\ mode of a diode laser is filtered spatially using a pinhole
(PH), two lenses and an iris. \ Light is placed in the proper coherent
superposition of each of the rails labelled A, B, and C using half-wave
plates (}$\lambda $\textbf{/2) and polarizing beam-splitters (PBSs). \ Glass
plates (GPs) in each arm are used to displace each beam transverse to its
direction of propagation, and microscope slides (MS) are used to finely
adjust the phases in arms A and C. \ The three modes are recombined at two
50/50 beamsplitters (BS3 and BS4). \ The beam shines on a screen and a CMOS
camera behind that screen captures its image. \ A photodiode (PD) in the
second port of BS4 is used to set the relative phases light in the different
rails.}

\textbf{Figure 2. \ Sample experimental data when only rail C is displaced.
\ We show experimental data for individual horizontal beam profiles from
Rails A (thin solid line), B(thin dashed line), and C (thick dashed line)
when a transverse, horizontal displacement of beam C is applied. \ The beam
profile of the post-selected beam is also shown (thick solid line). \ The
beam profile from rail C is displaced by }$-11.1$\textbf{\ pixels, or }$%
-0.69 $\textbf{\ rms widths, which constitutes a measure of the coupling
strength }$K_{C}$\textbf{. \ The corresponding shift }$\Delta x$ \textbf{in
the post-selected beam\ is }$+7.0$\textbf{\ pixels, or }$+0.44$\textbf{\ rms
widths, implying a probability of }$P_{C}=-0.64$\textbf{. \ (This is smaller
than the expected }$P_{CW}=-1$\textbf{\ because in this image, the coupling
strength is too large for a true weak measurement. \ This image is shown for
clarity, while Figure 3 summarizes the data for weaker couplings as well,
where the agreement with weak-measurement theory is good.) \ }

\textbf{Figure 3. \ Experimental data for individual rail displacements. \
The displacement of the post-selected state is shown as a function of the
displacement of each individual rail. \ The data for rails A, B, and C are
shown as open circles, solid triangles, and solid circles respectively. \
Theory (with no adjustable parameters) is shown as a dashed line for rails A
and B and as a solid line for rail C. \ The transition for the displacement
of rail C from the weak measurement regime to the strong measurement regime
is clearly visible at a beam displacement of about }$\pm $\textbf{1 rms\
width. \ No such transition was observed or predicted for the other two
rails.}

\textbf{Figure 4. \ The weak probabilities for the photon to be in rail B
and rail C measured in the same experimental setup using two different
pointers. \ The weak probability for the photon to be in rail C was measured
using a small polarization rotation, and the probability for rail B was
measured with a transverse beam displacement. \ Both are shown as functions
of the displacement of rail B, with the polarization rotation being kept
fixed. \ The resulting polarization rotations and displacements in the
post-selected states were converted into weak probabilities by dividing by
the relevant interaction strengths. \ The resulting weak probabilities for
rails B and C are shown as solid circles and open triangles. \ For very weak
measurements, the entire post-selected state is polarization rotated and
displaced, by amounts which would imply that each photon was both in rail C
and rail B with certainty. \ However, when a polarizer is inserted to block
vertically-polarized light (such that we only detect the photons that have
been polarization-rotated in rail C) we find the weak probability for the
photon to have been in rail B to be zero (open circles, dashed line). \
Thus, those photons that were definitely in rail C were definitely not in
rail B.}


\begin{thebibliography}{99}
\bibitem{weakmeas} Y. Aharonov, D.Z. Albert, and L. Vaidman, \emph{Phys.
Rev. Lett.}, \textbf{60}, 1351 (1988).

\bibitem{leggett} A.J. Leggett, \emph{Phys. Rev. Lett.}, \textbf{62}, 2325
(1989).

\bibitem{peres} A. Peres, \emph{Phys. Rev. Lett.}, \textbf{62}, 2326 (1989).

\bibitem{reply} Y. Aharonov and L. Vaidman, \emph{Phys. Rev. Lett.}, \textbf{%
62}, 2327 (1989).

\bibitem{weakexp} N.W.M. Ritchie, J.G. Story, and R.G. Hulet, \emph{Phys.
Rev. Lett.}, \textbf{66}, 1107 (1991).

\bibitem{opticalweak} I.M. Duck, P.M. Stevenson, and E.C.G. Sudarshan, \emph{%
Phys. Rev. D}, \textbf{40}, 2112 (1989).

\bibitem{aetunnelweak} A.M. Steinberg, \emph{Phys. Rev. Lett.}, \textbf{74},
2405 (1995).

\bibitem{hardyvaid} Y. Aharonov, A. Botero, S. Popescu, B. Reznik, and J.
Tollaksen, \emph{Phys. Lett. A}, \textbf{301}, 130 (2001); L. Hardy, \emph{%
Phys. Rev. Lett.}, \textbf{68}, 2981 (1992).

\bibitem{wiseman} H.M. Wiseman, \emph{Phys. Lett. A}, \textbf{311}, 285
(2003).

\bibitem{molmer} K. M\o lmer, \emph{Phys. Lett. A}, \textbf{292}, 151 (2001).

\bibitem{orozco} G.T. Foster, L.A. Orozco, H.M. Castro-Beltran, and H.J.
Carmichael, \emph{Phys. Rev. Lett.}, \textbf{85}, 3149 (2000); H. M.
Wiseman, \emph{Phys. Rev. A}, \textbf{65}, 032111 (2002).

\bibitem{box} Y. Aharonov and L. Vaidman, \emph{J. Phys. A: Math. Gen.}, 
\textbf{24}, 2315 (1991).

\bibitem{weakelements} L. Vaidman, \emph{Foundations of Physics}, \textbf{26}%
, 895 (1996).

\bibitem{negprobs} R.P. Feynman, \emph{International Journal of Theoretical
Physics}, \textbf{21}, 467 (1982); M.O. Scully, H. Walther, W. Schleich, 
\emph{Phys. Rev. A}, \textbf{49}, 1562 (1994).

\bibitem{muckenheim} W. Muckenheim, \emph{Phys. Rep.}, \textbf{133}, 339
(1983).

\bibitem{ABL} Y. Aharonov, P.G. Bergmann, and J.L. Lebowitz, \emph{Phys. Rev.%
}, \textbf{134}, B1410 (1964).

\bibitem{fallingtree} A.M. Steinberg, ``Can a falling tree make a noise in
two forests at the same time?'', in \emph{Causality and Locality in Modern
Physics}, S. Jeffers, G. Hunter, and J.-P. Vigier, eds., Kluwer Academic
Publishers (Dordrecht: 1997), p. 431; A. M. Steinberg, S. Myrskog, Han Seb
Moon, Hyun Ah Kim, Jalani Fox, Jung Bog Kim, \emph{Ann. Phys. (Leipzig)}, 
\textbf{7}, 593 (1998).

\bibitem{aewheeler} A.M. Steinberg, to appear in SCIENCE AND ULTIMATE
REALITY: Quantum Theory, Cosmology and Complexity, eds. John D. Barrow, Paul
C.W. Davies, and Charles L. Harper, Jr., Cambridge University Press, 2003;
quant-ph/0302003.

\bibitem{jwv} K.J. Resch and A.M. Steinberg, ``Extracting joint weak values
with local, single-particle measurements'', submitted (2003).

\pagebreak
\end{thebibliography}
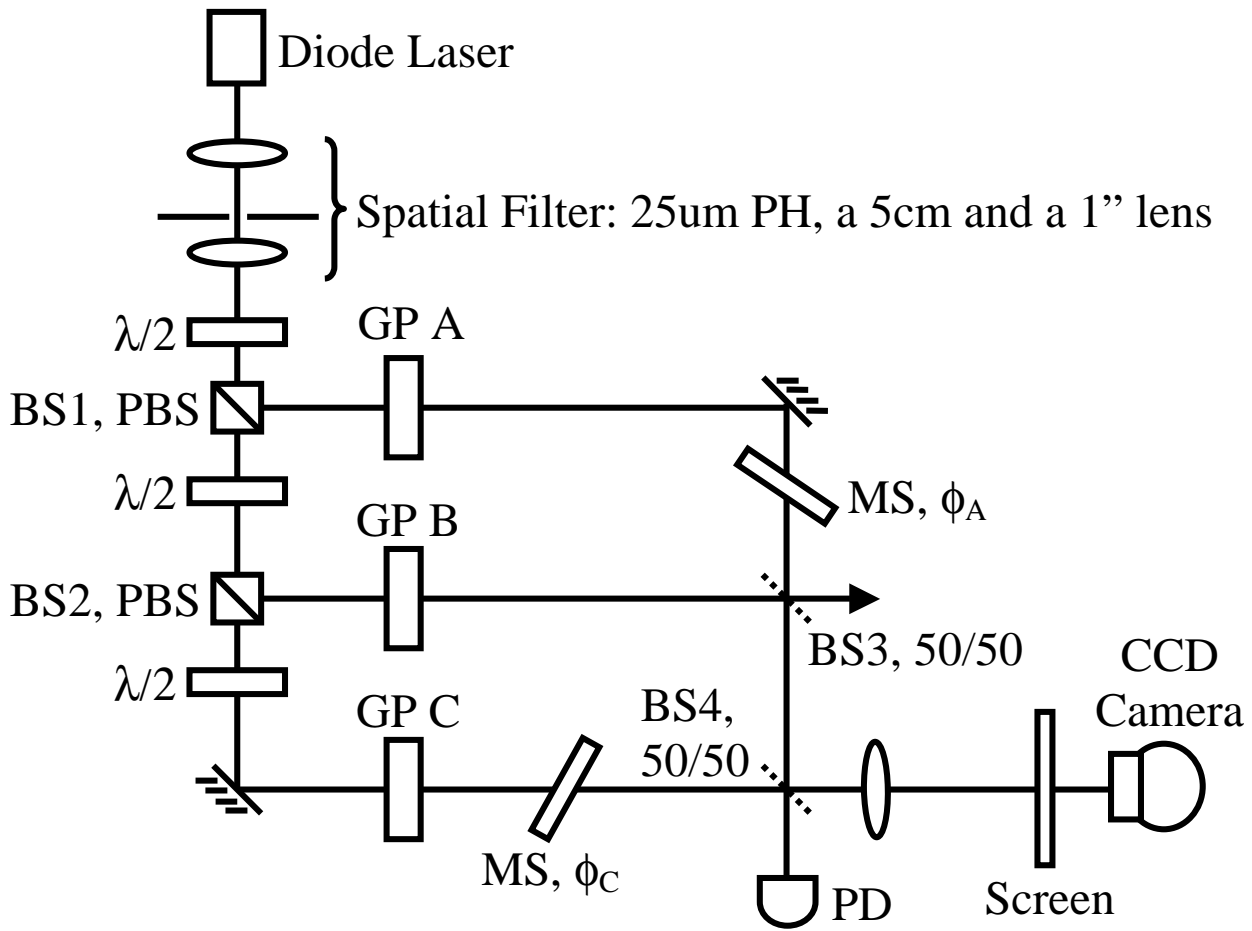
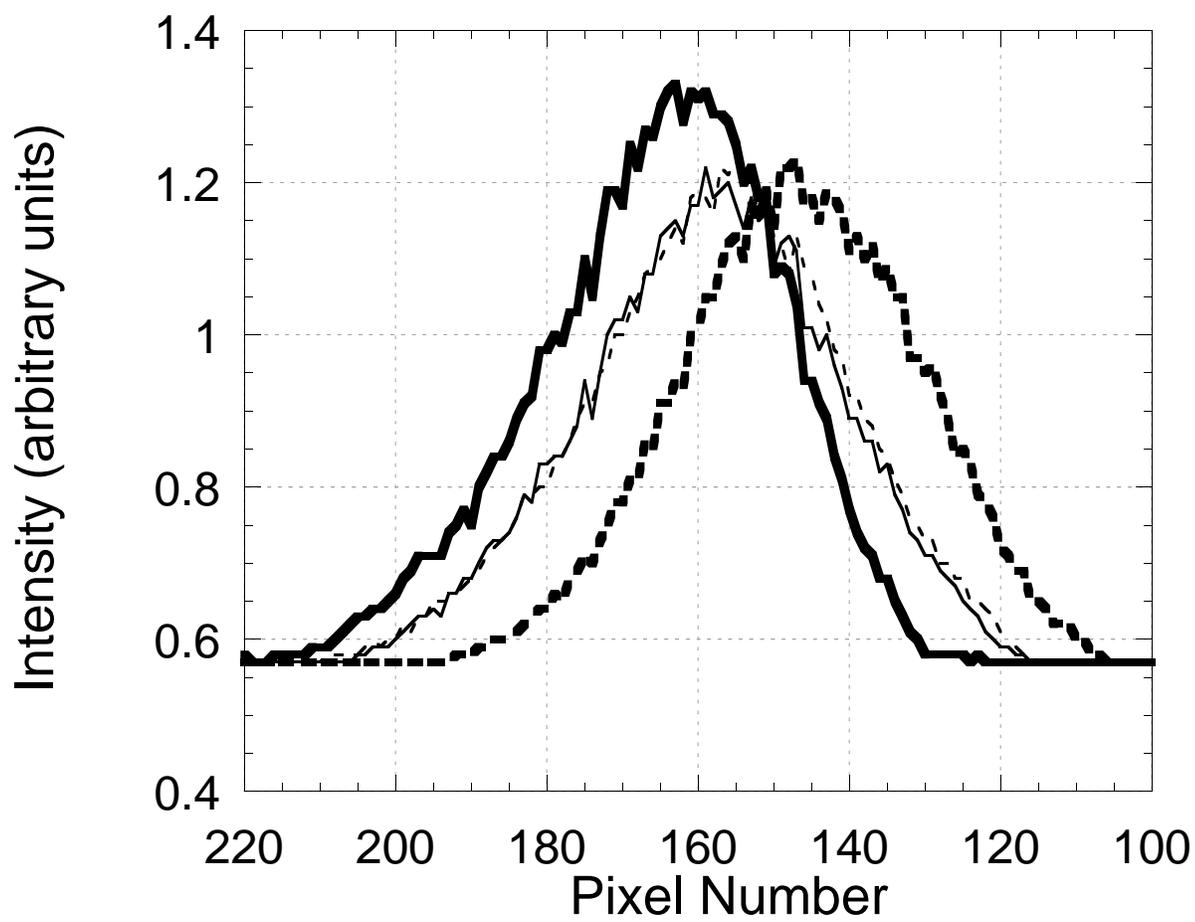
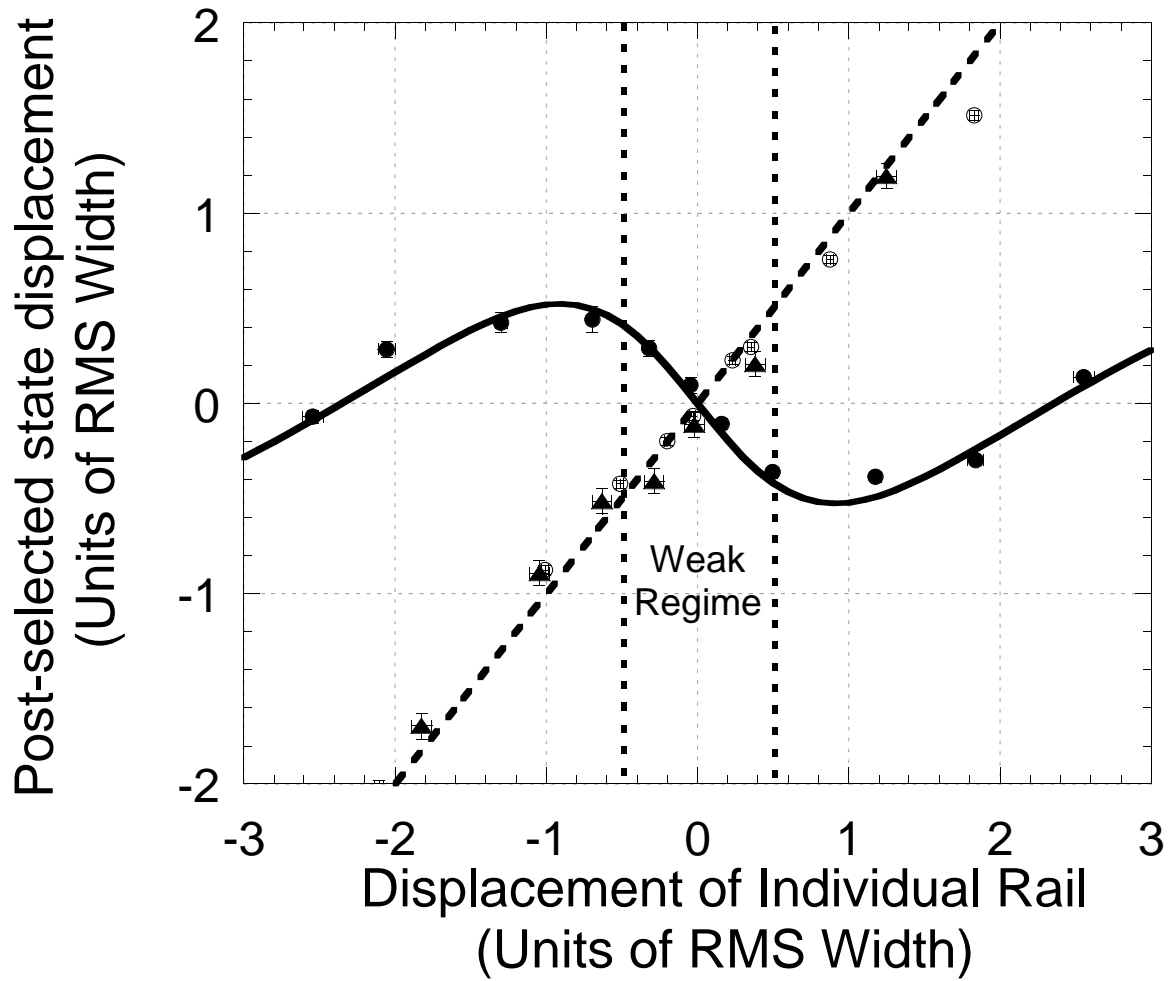
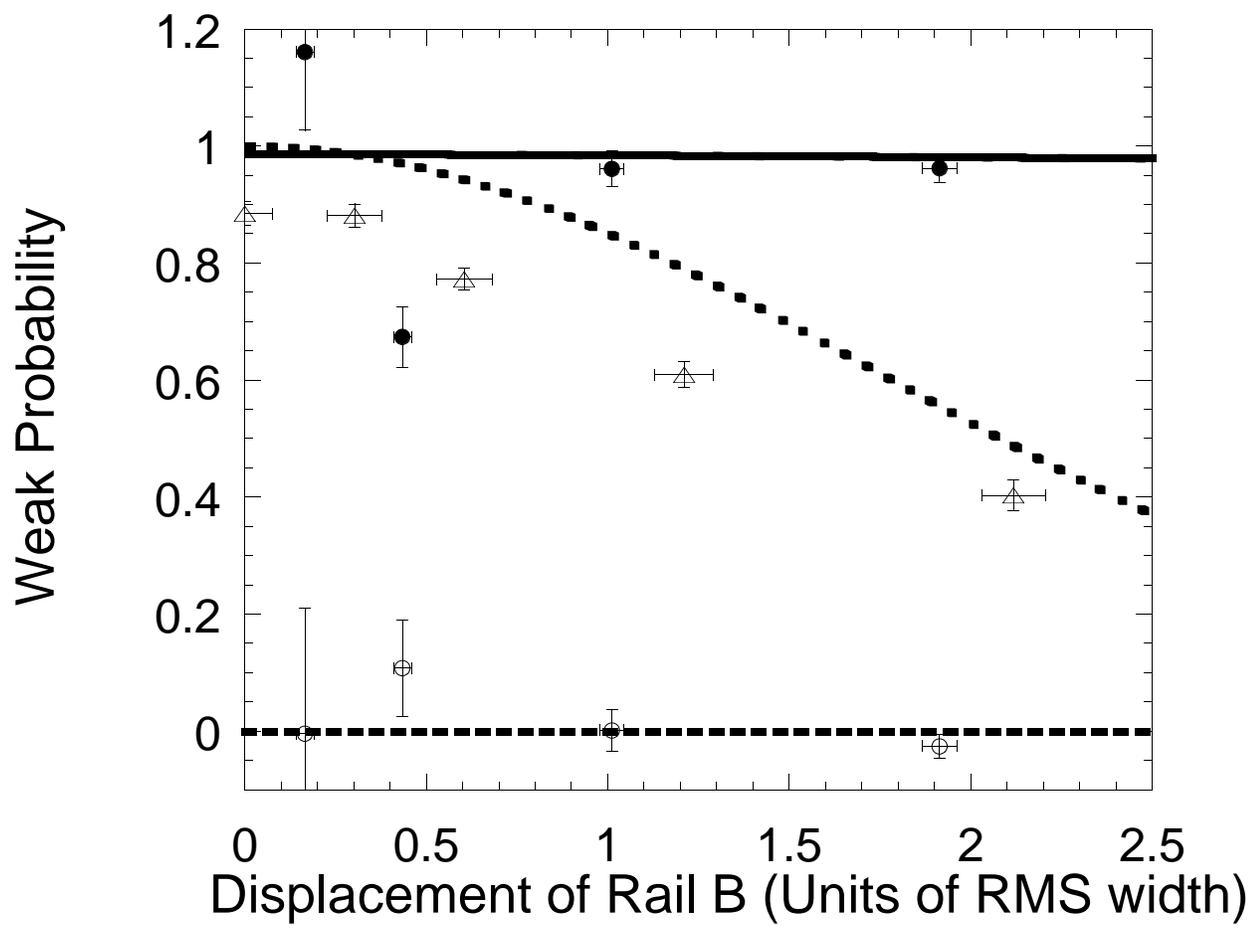
\end{document}